\documentclass[pra,showpacs,showkeys,amsmath,amssymb,superscriptaddress,preprintnumbers,floatfix, twocolumn]{revtex4}

\usepackage{bm,amsfonts, mathtools}
\usepackage{graphicx,color, wasysym}
\usepackage{textcomp}
\usepackage{hyperref}

\def\de{\mathrm{d}}

\begin{document}

\title{Operator solutions for fractional Fokker-Planck equations}

\author{K.~G\'{o}rska}
\email{kasia\_gorska@o2.pl}

\affiliation{Laboratoire de Physique Th\'{e}orique de la Mati\`{e}re Condens\'{e}e (LPTMC), Universit\'{e} Pierre et Marie Curie, CNRS UMR 7600, Tour 13 - 5i\`{e}me \'{e}t., Bo\^{i}te Courrier 121, 4 place Jussieu, F 75252 Paris Cedex 05, France}

\author{K.~A.~Penson}
\email{penson@lptl.jussieu.fr}

\affiliation{Laboratoire de Physique Th\'{e}orique de la Mati\`{e}re Condens\'{e}e (LPTMC), Universit\'{e} Pierre et Marie Curie, CNRS UMR 7600, Tour 13 - 5i\`{e}me \'{e}t., Bo\^{i}te Courrier 121, 4 place Jussieu, F 75252 Paris Cedex 05, France}

\author{D.~Babusci}
\email{danilo.babusci@lnf.infn.it}

\affiliation{INFN - Laboratori Nazionali di Frascati, v. le E. Fermi, 40, I 00044 Frascati (Roma), Italy}

\author{G.~Dattoli}
\email{dattoli@frascati.enea.it}

\affiliation{ENEA - Centro Ricerche Frascati, v. le E. Fermi, 45, I 00044 Frascati (Roma), Italy}

\affiliation{Universit\'e Paris XIII, LIPN, Institut Galil\'{e}e, CNRS UMR 7030, 99 Av. J.-B. Clement, F 93430 Villetaneuse, France}

\author{G.~H.~E.~Duchamp}
\email{ghed@lipn-univ.paris13.fr}

\affiliation{Universit\'e Paris XIII, LIPN, Institut Galil\'{e}e, CNRS UMR 7030, 99 Av. J.-B. Clement, F 93430 Villetaneuse, France} 


\begin{abstract}
We obtain exact results for fractional equations of Fokker-Planck type using evolution operator method. We employ exact forms of one-sided L\'{e}vy stable distributions to generate a set of self-reproducing solutions. Explicit cases are reported and studied for various fractional order of derivatives,  different initial conditions, and for different versions of Fokker-Planck operators.
\end{abstract}

\maketitle

\section{Introduction}

Ordinary derivatives account for the variation of a given function with respect to a given variable. Fractional derivatives have a more subtle meaning. We use throughout the Euler's definition of the fractional derivative according to which the derivative of order $\alpha$ ($0<\alpha<1$) of a constant is indeed not zero, but $\partial_{x}^{\alpha}\, 1 \,=\, \frac{x^{-\alpha}}{\Gamma(1 - \alpha)}$ \cite{AAKilbas06}. Their role in modelling physical phenomena is not intuitive and the treatment of the associated fractional differential equations (FDE) requires extreme care, not only from the mathematical point of view. The generalization of a relaxation equation, with a constant force term, to a fractional form reads \cite{TFNonnenmacher95, BJWest10}
\begin{equation}\label{eq1}
\partial_{t}^{\alpha}\, P_{\alpha}(t) \,=\, -\kappa\, P_{\alpha}(t) + \frac{t^{-\alpha}}{\Gamma(1 - \alpha)}\, P + f,
\end{equation}
where $P = P_{\alpha}(0)$ is the initial condition and $\kappa$ is a constant. Eq.~(\ref{eq1}) is an $\alpha$th order FDE with $f$ being the  non-homogeneous part. The term with $P$ is not a genuine non-homogeneous contribution, but it accounts  for the nonvanishing of a constant under fractional derivative. We use in Eq.~(\ref{eq1}) the Euler definition of fractional derivative because it appears most suitable to treat the dynamical behavior governed by the FFP equation we will discuss later. The problem (\ref{eq1}) is mathematically well defined. The apparent singularity at $t=0$ can be removed by multiplying both sides of the equation by $\partial_{t}^{1-\alpha}$, thus getting
\begin{equation}\label{eq2}
\partial_{t}\, P_{\alpha}(t) \,=\, \partial_{t}^{1-\alpha} \left(- \kappa\, P_{\alpha}(t) + f\right).
\end{equation}
The notion of stationary solution is well defined for an ordinary relaxation differential equation, but not for its fractional counterpart. In common terms \textit{stationary} means that the solution is no more sensitive to time variations and, hence, its (ordinary) time derivative is zero. The notion should be revised for FDE, in accordance with the order of the derivative. The solution of Eq.~(\ref{eq1}) reads \cite{BJWest10, IPodlubny99}:
\begin{eqnarray}\label{eq3}
P_{\alpha}(t) = E_{\alpha}\left(-\kappa\, t^{\alpha}\right) + f\, t^{\alpha} E_{\alpha, \,\alpha + 1}\left(- \kappa\, t^{\alpha}\right),
\end{eqnarray}
where $E_{\alpha, \beta}(z) \,=\, \sum_{r = 0}^{\infty} z^{r}/\Gamma(\alpha\, r + \beta)$ is the modified Mittag-Leffler function and reduces to its ordinary case for $\beta = 1$, $E_{\alpha}(z) = E_{\alpha, 1}(z)$ \cite{AAKilbas06}.

\begin{figure}[!h]
\includegraphics[scale=0.4]{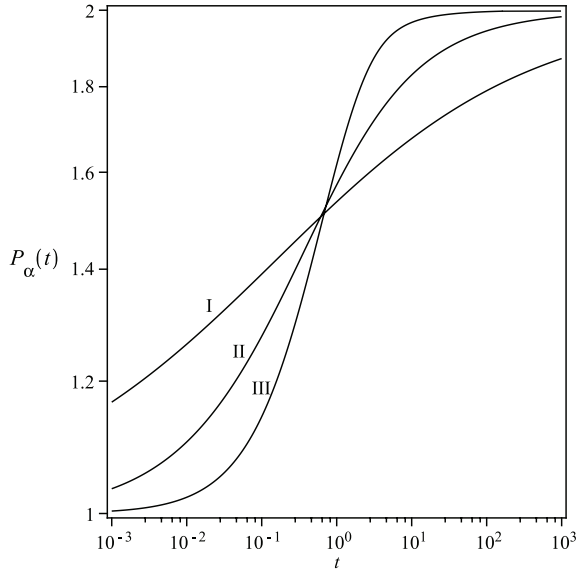}
\caption{\label{fig1} Logarithmic plot of $P_{\alpha}(t)$, (see Eq.~(\ref{eq3})), for $\kappa = 1$, $P = 1$, $f = 2$ and $\alpha =1/4$ (I), $1/2$ (II), and $5/6$ (III).}
\end{figure}

The solutions, plotted in Fig.~\ref{fig1} for different values of $\alpha$, do not display any long time stationary behavior. Quasi-stationary behavior is reached for $\alpha$ approaching the unity. For large $t$ we find $P_{\alpha}(t) \propto t^{\alpha - 1}$, for which the $\alpha$th derivative is vanishing. We can therefore conclude that the solution reaches $\alpha$-\textit{derivative stationary} form.

\section{Fractional Fokker-Planck equations and evolution operators}

The Eq.~(2) can be generalized to the following partial differential equation
\begin{equation}\label{eq4}
\partial_{t}^{\,\alpha}\, F_{\alpha}(x, t) = \hat{L}_{FP}\, F_{\alpha}(x, t) + \frac{t^{-\alpha}}{\Gamma(1-\alpha)}\, \gamma(x),
\end{equation}
which has been shown to be tailor suited for study of problems of anomalous diffusion  \cite{RMetzler99}. The initial condition is $F_{\alpha}(x, 0) = \gamma(x)$. From mathematical point of view Eq.~(\ref{eq4}) is a well-posed Cauchy problem and it is the two-variables generalization of Eq.~(\ref{eq1}). In Ref.~\cite{RMetzler99} Eq.~(\ref{eq4}) has been used to model the continuous time random walk with the inclusion of effects of space dependent jump probabilities and $\hat{L}_{FP}$ denotes the Fokker-Planck (FP) operator involving the spatial derivative $\partial_{x}$. The presence of the term with $\gamma(x)$ ensures that Eq.~(\ref{eq4}) is well defined and describes a process preserving the norm of the distribution $F_{\alpha}(x, t)$, when the time evolves. For any function $h(x)$ its average value over 
$F_{\alpha}(x, t)$ is defined as
\begin{equation}\label{eqhav} 
\langle h(t)\rangle_{\alpha} = \int_{-\infty}^{\infty} h(x)\, F_{\alpha}(x, t)\,\de x\,.
\end{equation}

The formal solution of Eq.~(\ref{eq4}) is obtained by using an extension of the evolution operator formalism, introduced by Schr\"{o}dinger, therefore getting
\begin{equation}\label{eq5}
F_{\alpha}(x, t) \,=\, \hat{U}_{\alpha}(t)\, \gamma(x), \qquad \hat{U}_{\alpha}(t) \,=\, E_{\alpha}(t^{\alpha}\, \hat{L}_{FP}).
\end{equation}

Below we shall apply Eqs.~(\ref{eq5}) to three different versions of Fokker-Planck operators $\hat{L}_{FP}$. Limiting for the moment the discussion to 
$\hat{L}_{FP} \,=\, k\, \partial_{x}^{2}$, where $k$ is the generalized diffusion constant, Eq.~(\ref{eq4}) can be interpreted as the fractional version of the heat equation \cite{DVWidder75} and its solution reads
\begin{equation}\label{eq6}
F_{\alpha}(x, t) \,=\, \sum_{r = 0}^{\infty} \frac{(k\, t^{\alpha})^{r}}{\Gamma(1 + \alpha\, r)}\, \left[\partial_{x}^{2\, r}\, \gamma(x)\right],
\end{equation}
which for $\gamma(x) = x^{n}$ ($n \in \mathbb{Z}$), gives
\begin{equation}\label{eq7}
F_{\alpha}(x, t) \,=\, _{\alpha}H_{n}^{(2)}(x, k\,t^{\alpha}) \,=\, n!\, \sum_{r = 0}^{[n/2]} \frac{x^{n-2\, r}\, (k\, t^{\alpha})^{r}}{(n - 2\, r)!\, \Gamma(1 + \alpha\, r)},
\end{equation}
which are polynomials in $x$. For $\alpha = 1$ they are known as heat polynomials \cite{DVWidder75}. Any initial function $\gamma(x)~=~\sum_{n=0}^{\infty} c_{n}\, x^{n}$ allows therefore 
a solution of the fractional heat equation as the following expansion
\begin{equation}\label{eq8}
F_{\alpha}(x, t) \,=\, \sum_{n=0}^{\infty} c_{n}\, _{\alpha}H_{n}^{(2)}(x, k\,t^{\alpha}).
\end{equation}

As in the case of conventional heat equation, the series in terms of \textit{fractional} heat polynomials $_{\alpha}H^{(2)}_{n}$ are of limited usefulness since it converges for short times only.  As an example, for $\alpha=1$ and 
\begin{equation}\label{eqfx}
\gamma(x) = \frac{1}{\sqrt{2\pi}\, \sigma_{x}}\, e^{-x^{2}/(2\, \sigma_{x}^{2})}\,,
\end{equation}
the convergence is limited to $t < \sigma_{x}^{2}/(4\, k)$. The use of the Gauss-Weierstrass transform \cite{APPrudnikov92} provides solutions with a well behaved long time behavior and therefore we look for an analogous transform relevant for the fractional case.

We make therefore the assumption that such a transform exists and that we can write
\begin{equation}\label{eq9}
E_{\alpha}(a\, t^{\alpha}) \,=\, \int_{0}^{\infty} n_{\alpha}(s, t)\, e^{a\, s}\, \de s
\end{equation}
with $n_{\alpha}(s, t)$ being not yet specified functions. The evolution operator in Eq.~(\ref{eq5}) can therefore be written as
\begin{equation}\label{eq12}
\hat{U}_{\alpha}(t) \,=\, \int_{0}^{\infty} n_{\alpha}(s, t)\, \hat{U}_{1}(s)\, \de s, \qquad \hat{U}_{1}(t) \,=\, e^{t\, \hat{L}_{FP}},
\end{equation}
and, equivalently, 
\begin{equation}\label{eq12a}
F_{\alpha}(x, t) \,=\, \int_{0}^{\infty} n_{\alpha}(s, t)\, F_{1}(x, s)\, \de s
\end{equation}
holds. Eq.~({\ref{eq12a}}) provides the link between fractional and ordinary Fokker-Planck equations through the knowledge of $n_{\alpha}(s, t)$. This equation, specified to the case of Eq. (\ref{eq7}), leads to the following relation
\begin{equation}\label{eq10a}
_{\alpha}H_{n}^{(2)}(x, t^{\alpha}) \,=\, \int_{0}^{\infty} n_{\alpha}(s, t)\, _{1}H_{n}^{(2)}(x, s)\, \de s,
\end{equation}
which yields, as a direct consequence of Eq.~(\ref{eq9})
\begin{equation}\label{eq10}
\int_{0}^{\infty} n_{\alpha}(s, t)\, \frac{s^{m}}{m!}\, \de s \,=\, \frac{t^{\alpha\, m}}{\Gamma(1 + \alpha\, m)}.
\end{equation}
According to Eq.~(16) in \cite{EBarkai01}, the functions $n_{\alpha}(s, t)$ can be identified as
\begin{equation}\label{eq11}
n_{\alpha}(s, t) \,=\, \frac{1}{\alpha}\, \frac{t}{s^{1 + 1/\alpha}}\, g_{\alpha}\left(\frac{t}{s^{1/\alpha}}\right).
\end{equation}
The functions $g_{\alpha}(z)$ are the one-sided L\'{e}vy stable distributions, recently obtained  for $\alpha$ rational  in \cite{KAPenson10, KGorska11}. For related considerations compare \cite{ASaa11}. The Eq.~(\ref{eq12}) is similar to the one reported in Refs.~\cite{RMetzler99} and \cite{EBarkai01}. The meaning of the fractional evolution operator is that the solution of the fractional Fokker-Planck (FFP) equation of order $\alpha$ is known whenever that the ordinary case, $\alpha = 1$, is available. By simple manipulation of the previous equations (see Eqs.~(\ref{eq9}) and (\ref{eq12})) we can also conclude that
\begin{equation}\label{eq15}
\hat{U}_{\beta}(t) \,=\, \int_{0}^{\infty}\, n_{\beta/\alpha}(s, t)\, \hat{U}_{\alpha}(s)\, \de s, \qquad \beta < \alpha\, .
\end{equation} 
Therefore, the solution of the FFP equation of order $\beta$ can be derived from its $\alpha$ counterpart by a self-reproducing procedure.  It should also be noted that, for 
$\alpha \neq 1$, $\hat{U}_{\alpha}(t_{2} + t_{1}) \neq \hat{U}_{\alpha}(t_{2})\, \hat{U}(t_{1})$. The evolution at different times $t_{2} > t_{1}$ is therefore given by
\begin{eqnarray}\label{eq12b}
\hat{U}_{\alpha}(t_{2}) \hat{U}_{\alpha}(t_{1}) &=& \int_{0}^{\infty} n_{\alpha}(s_{1}, t_{1})\, \de s_{1} \\ [0.7\baselineskip] \nonumber
&& \,\times\,  \int_{0}^{\infty}  n_{\alpha}(s_{2}, t_{2}) \hat{U}_{1}(s_{1} + s_{2}) \de s_{2}\,.
\end{eqnarray}
This formula turns out extremely useful to deal with successive approximations, when the nature of the FP operator does not provide any close form for the operator $\hat{U}_{1}(t)$. The functions $n_{\alpha}(x, t)$ defined in Eq.~(\ref{eq11}) turn out to be, for $\alpha = 1/k$, $k=2, 3, \ldots$, solutions of general heat equations $\partial_{t}\, u_{1/k}(x, t) = (-1)^{k}\, \partial^{2}_{x}\, u_{1/k}(x, t)$ with the initial condition $u_{1/k}(x, 0) = \delta(x)$. These heat equations have been also obtained in \cite{EOrsingher09} from purely probabilistic arguments. The case of $u_{\alpha}(x, t)$ for $\alpha = l/k$ and $l > 1$ will be the subject of a forthcoming study. 

\section{Specific examples}

We can now apply the wealth of the operator techniques known for the conventional FP equation, to solve its fractional version. For instance, starting with Gaussian initial condition of Eq. (\ref{eqfx}), we evaluate $\hat{U}_1 (s)\, \gamma(x)$ with the Glaisher formula \cite{GDattoli97, GDattoli00} and obtain 
\begin{eqnarray}\label{eq12c}
\hat{U}_{1}(s) \gamma(x) &=& \frac{1}{\sqrt{2\pi}\, \sigma_{x}}\, \left(1 + \frac{2 \kappa_1 s}{\sigma_{x}^2}\right)^{-1/2} \nonumber \\ 
& & \,\times \, \exp\left[-\frac{x^2}{2 \sigma_{x}^2} \left(1 + \frac{2 \kappa_1 s}{\sigma_{x}^2}\right)^{-1}\right],
\end{eqnarray}
which, according to formula (\ref{eqhav}), gives $\langle x^2(s)\rangle_{1} = \sigma_{x}^{2}\left(1 + \frac{2 k s}{\sigma_{x}^{2}}\right)$, and, by using Eq. (\ref{eq11}), allows us to conclude that the $\alpha$- and $t$-dependent variance of $x$ is given by
\begin{eqnarray}\label{eq12d}
\sigma_{x, \alpha}^{2}(t) &=& \int_{0}^{\infty} n_{\alpha}(s, t) \langle x^{2}(s) \rangle_{1}\, \de s \\[0.7\baselineskip] \nonumber
&=& \sigma_{x}^{2} + \frac{2 k t^{\alpha}}{\Gamma(1 + \alpha)}.
\end{eqnarray}
Note that for $\gamma(x) = \delta(x)$, we have formally $\sigma_{x}^{2} = 0$ and Eq.~(\ref{eq12d}) reproduces the defining equation of subdiffusive behavior. 

\begin{figure}[!ht]
\includegraphics[scale=0.86]{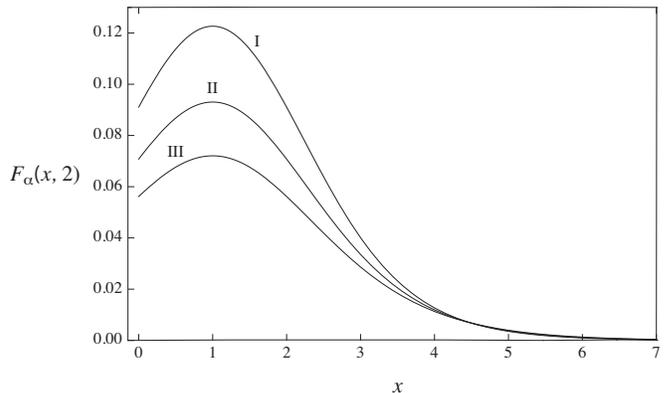}
\caption{\label{fig2} Plot of the solution of Eq.~(\ref{eq4}), $F_{\alpha}(x, t)$, with FP operator $\hat{L}_{FP} = \partial_{x}^{2} + 2\partial_{x}$ and initial condition 
$\gamma(x) = \frac{1}{\sqrt{2 \pi}}\, e^{-x^2/2}$, for $t=2$ and $\alpha =1/4$ (I), $1/2$ (II), and $3/4$ (III).}
\end{figure}

We now move on to more general Fokker-Planck operators. We start by considering the operator $\hat{L}_{FP}~=~k \partial_{x}^{2}~+~[\mathcal{F}/(m_{0}\, \eta_{\alpha})]\, \partial_{x}$, where the second term is due to the action of a constant force $\mathcal{F}$, $\eta_{\alpha}$ is fractional friction coefficient, and $m_{0}$ is the particle mass. The solution of our problem can be written by adding to the Glaisher form a shift term in the Gaussian provided by $\mathcal{F}\, t/(m_{0}\, \eta_{\alpha})$. The solution for different values of $\alpha$ and $t = 2$ are given in Fig. \ref{fig2} and the first moment of the distribution is, see Refs.~\cite{RMetzler99} and \cite{EBarkai01}: 
\begin{equation}\label{eq16}
\langle x(t) \rangle_{\alpha} \,=\, - \frac{\mathcal{F}\, t^{\alpha}}{m_{0}\, \eta_{\alpha}\, \Gamma(1 + \alpha)}.
\end{equation}    

The operator $\hat{L}_{FP} = \frac{2}{\tau}\left(\sigma^2_\epsilon \,\partial_{x}^{2} + \partial_{x} x\right)$ is used in storage ring physics to model the effect of diffusion and damping ($\tau$ is the damping time) of the electron beam due to the synchrotron radiation emitted by the electron in the bending magnets of the ring \cite{FCiocci00}. $\sigma_{\epsilon}$ is the variance of so-called \textit{equilibrium} distribution (see below). The two processes, (damping and diffusion), yield eventually a stationary solution in the conventional case. Such a condition does not exist for the fractional version. 

The evolution operator $\hat{U}_{1}(t)$ associated with the last FP operator can be written in a simple form. By setting indeed  $\hat{A} = 2\frac{t}{\tau} \sigma^2_\epsilon \partial_{x}^{2}$ and $\hat{B} = 2\frac{t}{\tau}\partial_{x} x$ we obtain $\left[\hat{A}, \hat{B}\right] = \frac{4 t}{\tau} \sigma^2_\epsilon \hat{A}$, so that the use of conventional operator ordering methods yields \cite{GDattoli97}
\begin{equation}\label{eq17}
\hat{U}_{1}(t) \,=\, e^{\hat{A} + \hat{B}} = \exp\left(\frac{1-e^{-4 t/\tau}}{4 t/\tau}\, \, \hat{A}\right)\, e^{\hat{B}}\,.
\end{equation}
In the case in which the initial distribution is the Gaussian the fractional evolution will be characterized by the following variance
\begin{equation}\label{eq18}
\langle x^{2}(t)\rangle_{\alpha} \,=\, \int_{0}^{\infty} n_{\alpha}(s, t)\, \langle x^{2}(s)\rangle_{1}\, \de s
\end{equation}
with 
\begin{equation}\label{eqseps}
\langle x^{2}(t)\rangle_{1} \,=\, \left(\sigma^{2} - \sigma_{\epsilon}^{2}\right)\, e^{-4 t/\tau} + \sigma_{\epsilon}^{2},
\end{equation}
where $\sigma$ is the variance of the initial distribution. In Eq.~(\ref{eq18}) $\langle x^{2}(t) \rangle_{\alpha}$ is obtained by replacing, according to Eq.~(\ref{eq9}) in the second term $e^{-4 t/\tau}$ with $E_{\alpha}\left(- 4 t^{\alpha}/\tau\right)$. (Note that the physical dimension of the damping time $\tau$ is $\left[t^{\alpha}\right]$). In Fig.~\ref{eq3} we reported the $\langle x^{2}(t)\rangle_{\alpha}$ and, as expected, equilibrium conditions in the conventional sense is not reached. The plot shows however the onset of analogous regimes after the knee-shaped decrease. This is a consequence of the fact that for increasing time the second term in Eq. (\ref{eqseps}) becomes dominating with respect to the first, and all solutions approach the 
$\alpha$-derivative stationary form.
\begin{figure}[!h]
\includegraphics[scale=0.35]{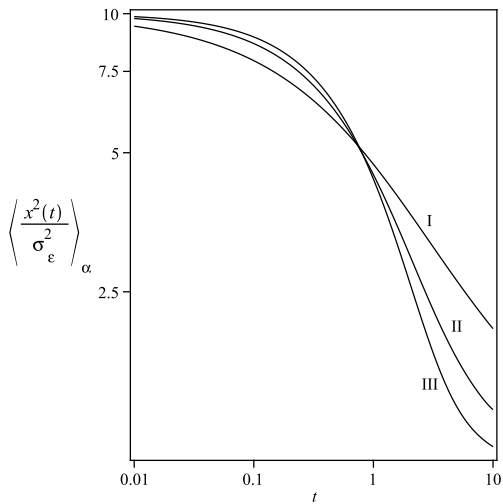}
\caption{\label{fig3} Double logarithmic plot of $\langle x^{2}(t)\rangle_{\alpha}$, (see Eq.~(\ref{eq18})), for $\sigma = 2$, $\sigma_{\epsilon} = 1$, and $\alpha = 3/5$ (I), $4/5$ (II), and $9/10$ (III).}
\end{figure}

\section{Discussion and conclusion}

We can also consider the case of partial fractional differential equations in which the fractional derivatives is acting on the \textit{spatial} coordinates. From the mathematical point of view we have the following Cauchy problem
\begin{eqnarray}\label{eq19}
\partial_{t}\,G_{\alpha}(x, t) &=& - \lambda\, \partial_{x}^{\alpha}\, G_{\alpha}(x, t) \\ [0.7\baselineskip] \nonumber
G_{\alpha}(x, 0) &=& h(x),
\end{eqnarray}
In such a context the Levy stable distribution function is going to play a role in the theory of FFP of type Eq.~(\ref{eq19}). The use of the evolution operator yields a formal solution of the type $G_{\alpha}(x, t) = e^{- \lambda\, t\, \partial_{x}^{\alpha}}\, h(x)$. The series expansion of the exponential may have a limited use only, we look therefore for a more useful representation of the evolution operator. The use of the identity \cite{KAPenson10}
\begin{equation}
e^{- a p^{\alpha}} = \int_{0}^{\infty} g_{\alpha}(\xi)\, \exp(- a^{1/\alpha}\, p\, \xi)\, \de \xi  
\end{equation}
is the naturally suited choice, so that we find
\begin{equation}\label{eq20}
G_{\alpha}(x, t) = \left(\lambda\, t\right)^{-1/\alpha}\, \int_{-\infty}^{x}\, g_{\alpha}\left[\frac{x-\sigma}{(\lambda\, t)^{1/\alpha}}\right]\, h(\sigma)\, \de \sigma\,.
\end{equation}
This technique (albeit limited to the case $\alpha = 1/2$) has been applied to the study of the relativistic heat equation 
($\partial_{t}\, G_{1/2}(x, t) = - \sqrt{1 - \partial_{x}^{2}}\, G_{1/2}(x, t)$) in \cite{DBabusci} and appears a very promising tool in further applications, possibly involving relativistic 
quantum mechanics.

Finally, we emphasize that the solutions of Eq.~(\ref{eq4}) for $1 < \alpha \leq 2$ can also be obtained with the help of the evolution operator and of \textit{two-sided} L\'{e}vy stable distributions obtained in \cite{KGorska11}. The form of Eq.~(\ref{eq9}) has to be however modified as then the $n_{\alpha}(s, t)$ function has to be replaced by its two-variable counterpart discussed in \cite{KGorska11}. In this context we refer to Eqs.~(5.21) and (5.22) of \cite{EOrsingher09} where the two-sided case is studied for the heat-type FP equation. 

The different topics touched on in this paper have shown that the combined use of techniques from various fields (including statistical mechanics, theory of fractional calculus, ordinary quantum mechanics, etc) offers the appropriate tool to study new phenomena in the theory of anomalous diffusion.

\section{Acknowledgment}

The authors acknowledge support from Agence Nationale de la Recherche (Paris, France) under Program PHYSCOMB No. ANR-08-BLAN-0243-2. G.~Dattoli thanks the University Paris XIII for financial support and kind hospitality.


\end{document}